\documentclass[letterpaper,twocolumn,10pt]{article}
\usepackage{usenix2019_v3}
\usepackage{tikz}
\usepackage{amsmath}

\usepackage{filecontents}
\usepackage{xspace}
\usepackage{xcolor}
\usepackage{multirow}
\usepackage{multicol}
\usepackage{subcaption}
\usepackage{titlesec}
\titlespacing*{\section}
{0pt}{1.5ex}{2.0ex}

\definecolor{darkerblue}{RGB}{0,63,202}
\definecolor{sabzseyedi}{RGB}{14,109,25}
\definecolor{LightGray}{RGB}{247,247,247}

\newcommand{\algo}{SourceFinder\xspace}
\newcommand{\github}{GitHub\xspace}
\newcommand{\totalrepo}{97K\xspace} 
\newcommand{\precision}{89\%\xspace} 
\newcommand{\recall}{86\%\xspace} 
\newcommand{\fOne}{87\%\xspace} 
\newcommand{\malrepo}{8644\xspace} 
\newcommand{\sourceNum}{7504\xspace}   

\newcommand{\malauthor}{malware authors\xspace}
\newcommand{\MalAuthors}{Malware authors\xspace}
\newcommand{\malHackersNo}{18\xspace}

\newcommand{\curDataSet}{MCur\xspace}  

\newcommand{\sourceThresh}{Source Percentage\xspace}   
\newcommand{\sourceThreshS}{SourceThresh\xspace}   

\newcommand{\miii}[1]{{\bf {\textcolor{blue}{MF:}}{\textcolor{red}{#1}}}}

\newcommand{\adark}[1]{{\textcolor{blue}{\bf AD:}}
{\textcolor{darkerblue}{\bf #1}}}

\newcommand{\eatreminders}{
\renewcommand{\adark}[1]{}
\renewcommand{\miii}[1]{}
}
 \eatreminders

\begin{document}


\title{\algo: Finding Malware Source-Code from Publicly Available Repositories}
\author{
{\rm Md Omar Faruk Rokon}\\
CS, UCR\\
mroko001@ucr.edu
\and
{\rm Risul Islam}\\
CS, UCR\\
risla002@ucr.edu
 \and
{\rm Ahmad Darki}\\
CS,UCR\\
adark001@ucr.edu
 \and 
{\rm Vagelis E. Papalexakis}\\
CS,UCR\\
epapalex@cs.ucr.edu
 \and
{\rm Michalis Faloutsos}\\ 
CS,UCR\\
michalis@cs.ucr.edu
} 

\maketitle


\begin{abstract}
	Where can we find malware source code? 
This question is motivated by a real need:
there is a dearth of malware source code, which impedes 
various types of security research. 
Our work is driven by the following insight: public archives, like \github,
have a surprising number of malware repositories.
Capitalizing on this opportunity,
we propose, \algo, a supervised-learning approach  to identify  repositories of malware source code efficiently. 
We evaluate and apply our approach using  97K repositories from \github. 
First, we show that our approach 
identifies malware repositories with  \precision precision and \recall recall
using a labeled dataset.
Second, we use \algo to identify \sourceNum
malware source code repositories, which arguably constitutes the largest malware source code database. 
Finally, we study the fundamental properties and trends of the 
malware repositories and their authors. 
The number of such repositories appears to be growing by an order of 
magnitude every 4 years,
and \malHackersNo malware authors seem to be
``professionals" with well-established online reputation.\miii{We could revisit key results later.}
We argue that our approach and our large repository of malware source code can 
be a catalyst for research studies,
which are  currently not possible.



\end{abstract}

\title{Our Catchy Title}

\section{Introduction}
\vspace{-0.15in}
\begin{figure}
    \centering
    \includegraphics[height=5.5cm,width=0.9\linewidth]{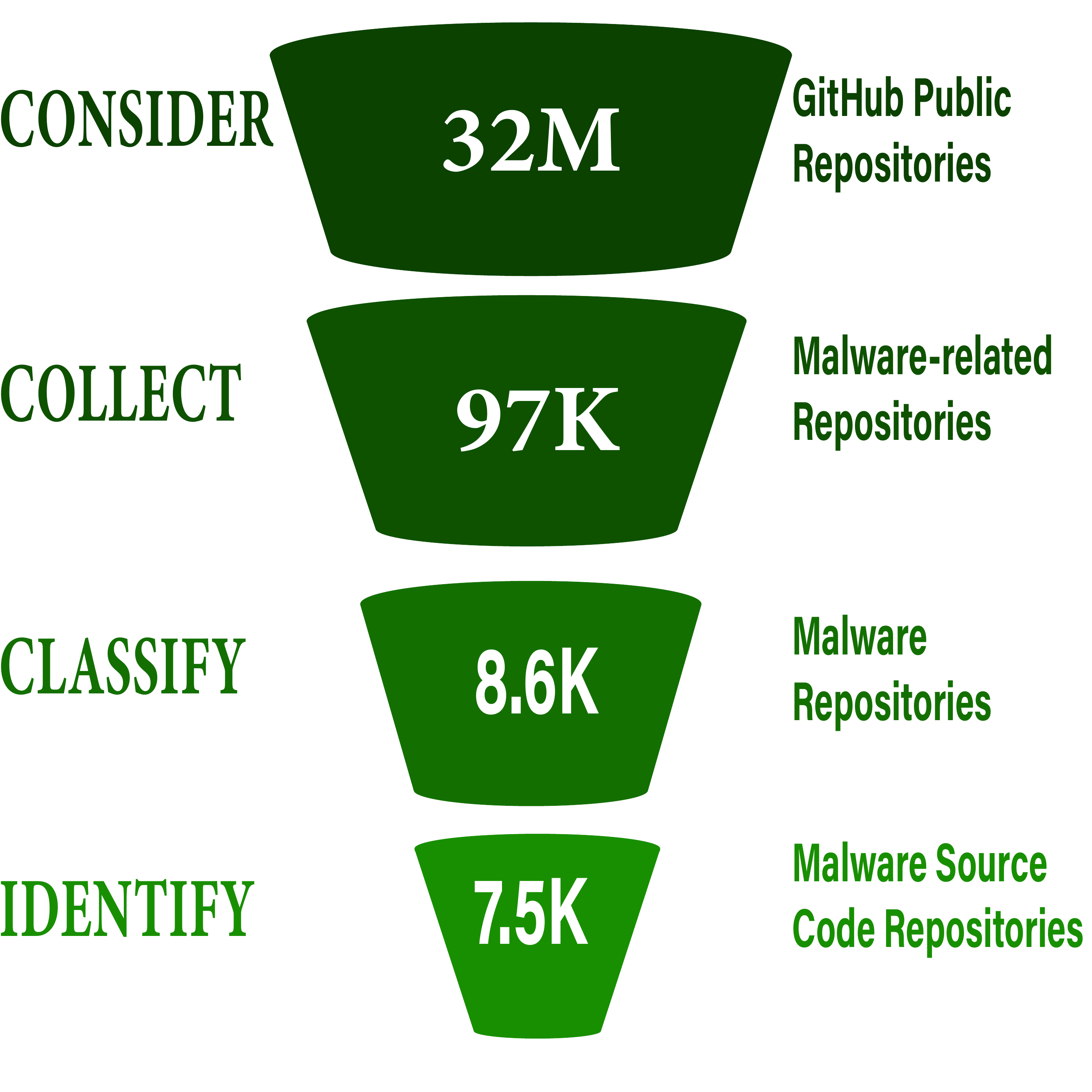}
    \caption{Starting from 32M \github repositories, we find 7.5K malware source code repositories using 137 malware keywords (Q137).}
    \label{fig:intro}
\end{figure}

Security research could greatly benefit by an extensive database of malware source code, which is currently unavailable.
This is the assertion that motivates this work.
First, security researchers can use malware source code
to: (a) understand malware behavior and techniques, 
and (b) evaluate security methods and tools~\cite{darki2019idapro,jerkins2017motivating}.
In the latter, having the source code can  provide the groundtruth for
assessing the effectiveness of different techniques, such as reverse engineering methods\cite{healey2019source, schulte2018evolving,chen11novel} and anti-virus methods.
Second, currently, a malware {\bf source code} database is not readily available.
By contrast, there are several databases with malware {\bf binary  code}, as collected via honeypots, but even those are often limited in number and not widely available. We discuss existing malware archives in Section~\ref{sec:related}. 

{\bf A missed opportunity:}
Surprisingly, software archives, like \github, host many publicly-accessible 
malware repositories, but this has not yet been explored to provide
security researchers with malware source code for their studies.
In this work, we focus on \github which is arguably  the largest software storing and sharing platform.
As of October 2019, \github 
reports more than 34 million users \cite{git:user} and more than 32 million public repositories~\cite{git:pub-repo}.
As we will see later, there are thousands of repositories that have malware source code, which seem to have escaped the radar of the research community so far.

Why do authors create public malware repositories? 
This question mystified us: these repositories expose
both the creators and the intelligence behind the malware.
Intrigued, we conducted a small investigation on malware authors, as we discuss below.





{\bf Problem:} How can we find malware source code repositories in a large archive, like \github? 
The input to the problem is an online archive and the 
desired output is a database of malware repositories.
The challenges include: (a) collecting an appropriate set of repositories from the potentially vast archive, (b) identifying the  repositories that contain malware.
Optionally, we also want to further help researchers that will potentially use these repositories, by determining additional properties, such as 
the most likely target platform, the malware type or family etc. 
Another practical challenge is the need to create the ground truth for validation purposes.






{\bf Related work:} 
To the best of our knowledge, there does not seem to be any studies 
focusing on the problem above. We group related works in the following categories.
First, several studies analyze software repositories 
to find usage and limitations without any focus on 
malware~\cite{cosentino2016findings}. 
Second,  several efforts maintain databases of malware binaries but without source code~\cite{allix2016androzoo,arp2014drebin}. 
Third, many efforts attempt to extract higher-level information from binaries,  such as lifting to Intermediate Representation(IR)~\cite{vdurfina2011design}, but it is really difficult to re-create the source code~\cite{chen2013refined}. In fact, such studies would benefit
from  our malware source-code archive to evaluate and improve their methods. Taking a software engineering angle, an interesting work~\cite{calleja2016look} compares the evolution of 150  malware source code repositories with that of benign software. 
We discuss related works in Section~\ref{sec:related}.

{\bf Contributions:} 
Our work is arguably the first to systematically identify
malware source code repositories from a massive public archive.
The contribution of this work is three-fold:
(a) we propose \algo, a systematic approach to 
identify  malware source-code repositories with high precision, 
(b) we create, arguably, the largest non-commercial malware source code archive with \sourceNum repositories,
and (c) we study patterns and trends of the repository ecosystem including temporal and author-centric properties and behaviors.
We apply and evaluate our method on the \github archive, though it could also be used on other archives, as we discuss in Section~\ref{sec:discussion}.




Our key results can be summarized in the following points, and some key numbers are shown in Figure~\ref{fig:intro}.

a. We collect {\bf \totalrepo malware-related repositories} from \github. In the collection, we overcome various practical limitations,
and we also generate an extensive  groundtruth  with 2013 repositories, as we explain in Section~\ref{sub-sec:data-collection}. 

b. {\bf \algo achieves \precision precision.}
We systematically consider different Machine Learning approaches, and carefully-created representations for the different 
fields of the repository, such as title, description etc.
We then systematically evaluate the effect of the different features, as we discuss in Section~\ref{sec:result}.
We show that we classify malware repositories with a \precision precision, \recall recall and \fOne F1-score using five fields from the repository. 
    
c. {\bf We identify \sourceNum malware source-code repositories}, which is arguably the largest malware source-code database in the research community. 
We have already downloaded the contents in these repositories, in case \github decides to deactivate them.
We also created a curated database of 250 malware repositories manually verified and spanning a wide range of malware types.
Naturally, we intend to make our datasets available for research purposes.


d. {\bf The number of new malware repositories in our data more than triples every four years}. The increasing trend is interesting and alarming at the same time.

e. {\bf We identify popular and influential repositories.} We identify the  malware repositories using three metrics of popularity: the number of watchers, forks and stars. 
We find 8 repositories that dominate the top-5 lists of all three metrics. 

f. {\bf We identify prolific and influential authors.}  
We find that 3\% of the authors have more than 300 followers.
We also find that 
 0.2\% of the authors have more than 7 malware repositories, with 
 the most prolific author {\em cyberthreats} having created 336  repositories.

g. {\bf We identify and profile \malHackersNo professional hackers.} We find \malHackersNo authors of malware repositories, who seem to have created a brand around their activities, as they use the same user names in security forums.
For example,  user {\em 3vilp4wn} (pronounced evil-pawn) is the author of a keylogger malware in 
\github, which the author is promoting in  
the {\em Hack This Site} forum using the same username.
We present our study of malware authors in Section~\ref{sec:authors}.


{\bf Open-sourcing for maximal impact: creating an engaged community.}
We intend to make our datasets and our tools available for research purposes.
Our vision is to create community-driven reference center,
which will provide:
(a) malware source code repositories,
(b) community-vetted labels and feedback,
and
(c) open-source tools for collecting and analyzing malware repositories.
Our goal is to expand our database
with more software archives. Although authors could  start hiding their repositories (see Section~\ref{sec:discussion}),
we argue that our already-retrieved database could have significant impact in enabling certain types of research studies.

\section{Background}
\label{backround-data}
\label{sub-sec:background}
\vspace{-0.15in}
We provide  background information on \github and the type of information that repositories have.


\github is  a massive world-wide software archive, which
enables users to share code through its public repositories
thus creating a global social network of interaction. For instance,
 first, users can collaborate on a repository. Second, users often "fork" projects: they copy and evolve projects.
Third, users can follow  projects,
and "up-vote" projects using "stars"  (think Facebook likes).
Although \github has many private repositories, there are 32 million public software repositories.

We  describe  the key elements of a \github repository. A repository  is equivalent to a project folder, and typically, each repository corresponds to a single software project.

A repository in \github has the following data fields:
a) title, b) description, c) topics, d) README file, e) file and folders, f) date of creation and last modified, g) forks, h) watchers, i) stars, and j) followers and followings, which we explain below.

{\bf 1.  Repository title: }
The title is a mandatory field and it usually consists of less than 3 words.




{\bf 2.  Repository description: }
This is an optional field that describes the objective of the project
and it is usually 1-2 sentences long. 

{\bf 3.  Repository topics: }
An author can optionally provide topics for her repository, in the form of tags, for example, \textit{``linux, malware, malware-analysis, anti-virus"}.
Note that  97\% of the repositories in our dataset have less than 8 topics.

{\bf 4.  README file: }
As expected, the README file is a documentation and/or light manual for the repository. This field is optional and its size varies from
 one or two sentences to many paragraphs. 
 For example, we found that 17.48\% of the README files
 in our repositories are empty. 

{\bf 5.  File and folders: }
In a well-constructed software, the  file and folder names of the source code can provide useful information. 
For example, some malware repositories contain files or folders with
indicative names, such as ``malware", ''source code" or even specific malware types or names of specific malware, like {\em mirai}. 


{\bf 6.  Date of creation and last modification: }
\github maintains the date of creation and last modification of a repository.
We find malware repository created in 2008 are actively being modified by authors till present. 

7. {\bf Number of forks:} 
Users can fork a public repository: they can create a clone
of the project~\cite{jiang2017and}. An user can fork any public repository to change locally and contribute to the original project if the owner accepts the modification. 
The number of forks is an indication of the popularity and impact of a repository.
Note that the number of forks indicates the number of distinct users that have forked a repository.

{\bf 8.  Number of watchers:}
Watching a repository is equivalent to   ``following" in the social media language. A ``watcher" will get notifications, if there is any new activity in that project. The numbers of watchers
is an indication of the popularity of a repository~\cite{dabbish2012social}. 

{\bf 9.  Number of stars: }
A user can ``star" a repository, which is equivalent to the ``like" function in social media~\cite{begel2013social}, and places the repository in the users favorite group, but does not provide constant updates as with the ``watching" function.  

{\bf 10. Followers: }
Users can also follow other users' work. 
If A follows B, A will be added to B's followers and B will be added to A's following list. The number of followers is an indication of the popularity of a user~\cite{lee2013github}.




\section{Data Collection} \label{sub-sec:data-collection}
\vspace{-0.15in}
\begin{table}
    \centering
    \small
    \begin{tabular}
        {|p{0.1\linewidth}|p{0.65\linewidth}|r|}
        \hline
        {\bf Set} & {\bf Descriptions} & {\bf Size} \\ \hline
        Q1 & Query set =  \{"malware"\}  & 1  \\ \hline
        Q50 & Query with 50 keywords with  Q1$\subset$Q50 & 50  \\ \hline
        Q137 & Query with 137 keywords with  Q50$\subset$Q137 & 137\\ \hline \hline \hline
        RD1 &  Retrieved repositories from query Q1 & 2775  \\ \hline
        RD50 & Retrieved repositories   from query  Q50 & 14332  \\ \hline 
        RD137 & Retrieved repositories  from query  Q137 & 97375  \\ \hline \hline \hline
        LD1 & Labeled subset of RD1 dataset & 379  \\ \hline
        LD50 & Labeled subset of RD50 dataset & 755  \\ \hline
        LD137 & Labeled subset of RD137 dataset & 879  \\ \hline \hline \hline
        M1 & Malware source code repositories in RD1 & 680  \\ \hline
        M50 & Malware source code repositories in RD50 & 3096  \\ \hline
        M137 & Malware source code repositories in RD137 & 7504  
        \\ \hline \hline \hline 
        \curDataSet & Manually verified malware source code dataset & 250 \\ \hline 
    \end{tabular}
    \caption{Datasets, their relationships, and their size.}
    \label{tab:my_label}
\end{table}


The first step in our work is to collect repositories from \github that have a higher chance of being related to malware.
Extracting repositories at scale from \github hides several subtleties and challenges, which we discuss below.

Using the \github Search API, a user can query with a set of keywords  and obtain the most relevant repositories.
We describe briefly how we select appropriate keywords, retrieve related repositories from \github and how we establish our ground truth.



{\bf A. Selecting keywords for querying: }
In this step, we want to retrieve repositories from \github in a way that: (a) provides as many as possible malware repositories,
and (b) provides a wide coverage over different types of malware.
For this reason, we select keywords from three categories: (a) 
 malware and security related keywords, such as malware and virus,
  (b) malware type names, such as ransomware and keylogger, and (c) popular malware names, such as mirai. Due to space limitations, we will provide the full list of keywords in our website at publication time for repeatability purposes.

We define three sets of keywords that we use to query \github. The reason is that we want to assess the sensitivity of the number of keywords on the outcome. Specifically,  we use the following query sets: (a) the {\bf Q1 set}, which only contains  the keyword ``malware"; (b) the {\bf Q50 set}, which contains 50 keywords, and (c) the {\bf Q137 set} which contains 137 keywords. 
The Q137 keyword set is a super-set of Q50, and Q50 is a superset of Q1. 
As we will see below, using the query set Q137 provides wider coverage, and we recommend in practice. 
We use the other two to assess the sensitivity of the results in the initial set of keywords.
We list our datasets in Table~\ref{tab:my_label}.


{\bf B. Retrieving related repositories:}
Using the Search API, we query \github with our set of keywords.
Specifically, we query \github with every keyword in our set separately.
In an ideal world, this would have been enough to collect
all related repositories:
 a query with ``malware" (Q1) should return the many thousands related repositories, but this is not the case.

The search capability hides several subtleties and limitations.
 First, there is a limit of 1000 repositories that a single search can return: we get the top 1000 repositories ordered by relevancy to the query.
Second, the GitHub API allows 30 requests per minute for an authenticated user and 10 requests per minute for an unauthenticated user.

{\bf Bypassing the API limitations.}
We were able to find a work around for the first limitation by using  ranking option. Namely, a user can specify her preferred ranking order for the results based on: (a) best match, (b) most stars, (c) fewest stars, (d) most forks, (e) fewest forks, (f) most recently updated, and (g) the least recently updated order. By repeating a query with all these seven ranking options, we can maximize the number of distinct repositories that we get.
This way, for each keyword in our set, we search with these seven different ranking preferences to obtain a list of \github repositories. 




{\bf C. Collecting the repositories:} We download all the repositories
identified in our queries using PyGithub~\cite{PyGithub}, and we obtain three sets of repositories RD1, RD50 and RD137.
These retrieved datasets have the same "subset" relationship
that they query sets have: RD1 $\subset$ RD50 $\subset$ RD137.
Note that we remove pathological repositories, mainly repositories with no actual content, or  repositories "deleted" by \github. 
For each repository, we collect and store: (a) repository-specific information, (b) author-specific information, and (c) all the code within the repository.

As we see from Table~\ref{tab:my_label}, using more and specialized malware keywords 
returns significantly more repositories. Namely, searching with the keyword ``malware" does return 2775 repositories, but searching with the Q50 and Q137 returns 14332 and 97375 repositories respectively.




\begin{table}
    \centering
    \footnotesize
    \begin{tabular}
        {|p{0.28\linewidth}|p{0.28\linewidth}|p{0.28\linewidth}|}
        \hline
        {\bf Labeled Dataset}  & {\bf Malware Repo.} & {\bf Benign Repo.}  \\ \hline
        LD137  & 313 & 566 \\ \hline
        LD50  & 326 & 429  \\ \hline
        LD1  & 186 & 193  \\ \hline

    \end{tabular}
    \caption{Our groundtruth: labeled datasets for each of the three queries, for a total of 2013 repositories.}
    \label{tab:dat_def}
\end{table}

{\bf D. Establishing the groundtruth}:
As there was no available groundtruth, we needed to establish our own.
As this is a fairly technical task, we opted for 
 domain experts instead of Mechanical Turk users, as recommended by recent studies~\cite{gharibshah2020rest}. We use three computer scientists to manually label 1000 repositories, which we selected in a uniformly random fashion, from each of our dataset RD137 and RD50 and 600 repositories from RD1.
The judges were instructed to independently investigate every repository thoroughly. 

{\bf Ensuring the quality of the groundtruth.}
To increase the reliability of our groundtruth, we took the following measures. 
First, we asked judges to label a repository {\em only}, if they were certain that it is malicious or benign and distinct, and leave it unlabeled otherwise. We only kept the repositories for which the judges agreed unanimously. 
Second, duplicate repositories were removed via manual inspection,
and were excluded from the final labeled dataset to avoid overfitting.
It is worth noting that we only found very few duplicates in the order of 3-5 in each dataset with hundreds of repositories.

With this process, we establish three separate labeled datasets named LD137, LD50, and LD1 starting from the respective malware repositories from each of our queries, as shown in Table~\ref{tab:dat_def}. 
Although the labeled datasets are not 50-50, they are 
representing both classes reasonably well, so that a naive solution
that will label everything as one class, would perform poorly. By contrast, our approach performs sufficiently well, as we will see in Section~\ref{sec:result}.

  As there is no available dataset, we argue that we make a sufficient size dataset by manual effort.

\section{Overview of our Identification Approach} 
\label{mal-identification}
\vspace{-0.15in}



Here, we describe our supervised learning algorithm to identify the repositories that contain malware.


{\bf Step 1. Data preprocessing:}
As in any Natural Language Processing (NLP) method, we  start with some initial processing of the text to improve the effectiveness of the solution. We briefly outline three levels of processing functionality.

\textbf{a. Character level preprocessing:} We handle the character level ``noise" by removing  special characters, such as  punctuation and currency symbols,  and fix Unicode and other encoding issues.

\textbf{b. Word level preprocessing:} 
We eliminate or aggregate words following the best practices of Natural Language Processing~\cite{jivani2011comparative}.
First, we remove article  words and other words that don't carry significant meaning on their own. 
Second, we use a stemming technique
 to handle inflected words. Namely, we want to decrease the dimensionality of the data by grouping words with the same "root". For example, we group the words ``organizing'', ``organized'', ``organize'' and ``organizes'' to one word ``organize''.  
 Third, we filter out common file and folder names that we do not expect to help
 in our classification, such as ``LEGAL'', ``LICENSE'', ``gitattributes'' etc. 
 
\textbf{c. Entity level filtering:}  We filter entities that are likely not helpful in describing the scope of a repository. Specifically, we remove numbers, URLs, and emails, which are often found in the text.
We found that this filtering improved the classification performance.
In the future, we could consider mining URLs and other information, such as names of people, companies or youtube channels, to identify authors, verify intention, and find more malware activities.

{\bf Step 2. The repository fields:} 
We consider fields from the repositories that can be numbers or text.
Text-based fields require processing in order to turn them into 
classification features and we explain this below.
We use and evaluate the following text fields:  title, description, topics, file and folder names and README file fields.


\textbf{Text field representation: }
We consider two techniques to represent each text field by a feature in the classification.

{\bf i. Bag of Words (BoW):}
The bag-of-words (BoW) model  is among the most widely used representations of
a document. The document is represented as the  number of occurrences of its words, disregarding grammar and word order~\cite{zhang2010understanding}. This model is commonly used in document classification where the frequency of each word is used as feature value for training a classifier~\cite{mctear2016conversational}. We use the model with the count vectorizer and TF-IDF vectorizer to create the feature vector.

In more detail,
we represent each text field in the repository with a  vector $V[K]$, 
where $V[i]$ corresponds to the significance of work $i$ for the text.
There are several ways to assign values $V[i]$: (a) zero-one to account for presence, (b) number of occurrences, and (c) the TF-IDF value of the word. We evaluated all the above methods.

{\em Fixing the number of words per field.}
To improve the effectiveness of our approach using BoW, we conduct a feature selection process, ${\chi}^2$ statistic following best practices~\cite{rogati2002high}. The ${\chi}^2$ statistic measures the lack of independence between a word (feature) and a class. A feature with lower chi-square score is less informative for that class, and thus
not useful in the classification. 
We discuss this further in Section~\ref{sec:result}.
For each text-based field $f$,
we select the top $K_f$ words for that field, which exhibit the highest discerning power in identifying malware repositories.
Note that we set a value for $K_f$ during the training stage
For each field, we select  the value  $K_f$,
as we explain in Section~\ref{sec:result}.


{\bf ii. Word embedding:}
The word embedding model is a vector representations of each word in a document: each word is mapped to an M-dimensional vector of real numbers~\cite{mikolov2013distributed}, or equivalently are projected in an M-dimensional space. A good embedding  ensures that words that are close in meaning have nearby representations in the embedded space. 
In order to create the document vector, word embedding follows two approaches (i) frequency-based vectorizer(unsupervised)~\cite{schnabel2015evaluation} and (ii) content-based vectorizer(supervised)~\cite{kusner2015word}. 
Note that in this type of representation, we do not use the {\em word level processing}, which we described in the previous step, since this method can leverage contextual information. 

We use frequency-based word embedding with word average and TF-IDF vectorizer. We also use pre-trained model of Google word2vec~\cite{mikolov2013efficient} and Stanford (Glov)~\cite{pennington2014glove} to create the feature vector. 


Finally, we create the  vector of the repository by concatenating 
the vectors of each  field of that repository.


{\bf Step 3. Selecting the fields:} Another key question is which fields from the repository to use in our classification. We experiment with all of the fields listed in Section~\ref{sub-sec:background} and we explain our findings in the next Section.

{\bf Step 4. Selecting a ML engine:}
We design classifiers to classify the repositories into two classes: (i) malware repository and (ii) benign repository. 
We systematically evaluate many machine learning algorithms \cite{murphy2012machine,bishop2006pattern}: Naive Bayes (NB), Logistic Regression (LR), Decision Tree (CART), Random Forest(RF), K-Nearest Neighbor (KNN), Linear Discriminant Analysis (LDA), and Support Vector Machine (SVM). 



{\bf Step 5. Detecting source code repositories:}
We also want to identify the existence of source code in 
the repositories, as the final step in providing malware source code to the community.

We propose a heuristic approach, which seems to work fairly well in practice.
First, we want to identify files in the repository that contain source code.
To do this, we start by examining their file extension.
If the file extension is one of the known programming languages: {\em Assembly, C/C++, Batch File, Bash Shell Script, Power Shell script, Java, Python, C\#, Objective-C, Pascal, Visual Basic, Matlab, PHP, Javascript, and Go}, 
we label it as a source file. 
Second, if the number of source files in a repository exceeds 
the {\bf \sourceThresh threshold (\sourceThreshS)}, we consider 
that the repository contains source code.

{\em How effective is this heuristic?} It turns out that in practice it works pretty well, as we will see in Section \ref{sec:result}.
Given that authors go out of their way to share their malware openly, and even provide appropriate titles and keywords, it seems less likely that
they will attempt to obfuscate the existence of source code in the repository. 
\section{Evaluation: Choices and Results}
\label{sec:result}
\vspace{-0.15in}
In this section, we evaluate the effectiveness of the classification based on the proposed methodology defined in Section~\ref{mal-identification}. More specifically, our goal here is to answer the following questions:


\begin{enumerate}
\itemsep-0.3em
    \item {\bf Repository Field Selection:} Which repository fields should we consider in our analysis?
    \item {\bf Field Representation:} Which feature representation is better between bag of words (BoW) and word embeddings and considering several versions of each?
    \item {\bf Feature Selection}: 
    What are the most informative features in identifying malware repositories?
    \item {\bf ML Algorithm Selection:} Which ML algorithm exhibits the best performance?
    \item {\bf Classification Effectiveness:}
    What is the precision, recall and F1-score of the classification?
    \item {\bf Identifying Malware Repositories:} How many malware repositories do we find?
    \item {\bf Identifying Malware Source Code Repository: } How many of the malware repositories have source code? 
\end{enumerate}

Note that we have a fairly complex task: we want to identify the best fields, representation method and Machine Learning engine, while considering different values for parameters.
What complicates matters is that all these selections are interdependent. We present our analysis in sequence,
but we followed many  trial and error and non-linear paths in reality.

{\bf 1. Selecting repository fields:} 
We evaluated all the repository fields mentioned earlier.
In fact, we used a significant number of experiments with 
different subsets of the features, not shown here due to space limitations.
We find that the title, description, topics, README file, and file and folder names have the most discerning power.
 We also considered number of forks, watchers, and stars of the repository and the number of followers and followings of the author of the repository. We found that not only it did not help,
 but it usually decreased the classification accuracy by 2-3\%. 
 One possible explanation is that the numbers of forks, stars and followers reflect the popularity rather than the content of a repository. 

\begin{table}
    \centering
    \footnotesize
    \begin{tabular}
        {|p{0.65\linewidth}|p{0.22\linewidth}|}
        \hline
        {\bf Representation} & \textbf{Classification Accuracy Range}   \\ \hline
        Bag of Words with Count Vectorizer   & 86\%-51\% \\ \hline
        Bag of Words with Count Vectorizer + Feature Selection & 91\%-56\% \\ \hline
         Bag of Words with       TF-IDF vectorizer & 82\%-63\% \\ \hline \hline
        Word Embedding with Word Average  & 85\%-72\% \\ \hline 
        Word Embedding with TF-IDF  & 85\%-74\%  \\ \hline \hline 
        Pretrained Google word2vec Model & 76\%-64\% \\ \hline
        Pretrained Stanford (Glov) Model & 73\%-62\% \\ \hline
    \end{tabular}
    \caption{Selecting the feature representation model: We evaluate all the representations across seven machine learning approaches and report the range of the overall classification accuracy. }
    \label{tab:feature-representation}
\end{table}

{\bf 2. Selecting a field representation:}
The goal is to find, which representation approach works better.
In Table \ref{tab:feature-representation}, we show the comparison of the range of classification accuracy across the 7 different ML algorithms that we will also consider below.
We find that Bag of Words with the count vectorizer representation 
reaches 86\% classification accuracy,
with the word embedding approach nearly matches that with
85\% accuracy. 
Note that we finetune the selection of words to represent each field in the next step.

Why does not the embedding approach outperform the bag of words?
One would have expected that the most complex embedding approach would have been the winner and with a significant margin. 
We attribute this to the relatively small text size in most text fields, which also do not provide well-structured sentences (think two-three words for the title, and isolated words for the topics). Furthermore, the word co-occurrences does not exist in topics and file names field, which partly what makes embedding approaches work well in large and well structured documents \cite{li2015word, globerson2007euclidean}. 

In the rest of this paper, we choose the Bag of Words with count vectorizer to represent our text fields, since it exhibits good performance and is computationally less intensive than the embedding method.

\begin{figure}
    \centering
    \includegraphics[height=5cm, width=0.99\linewidth]{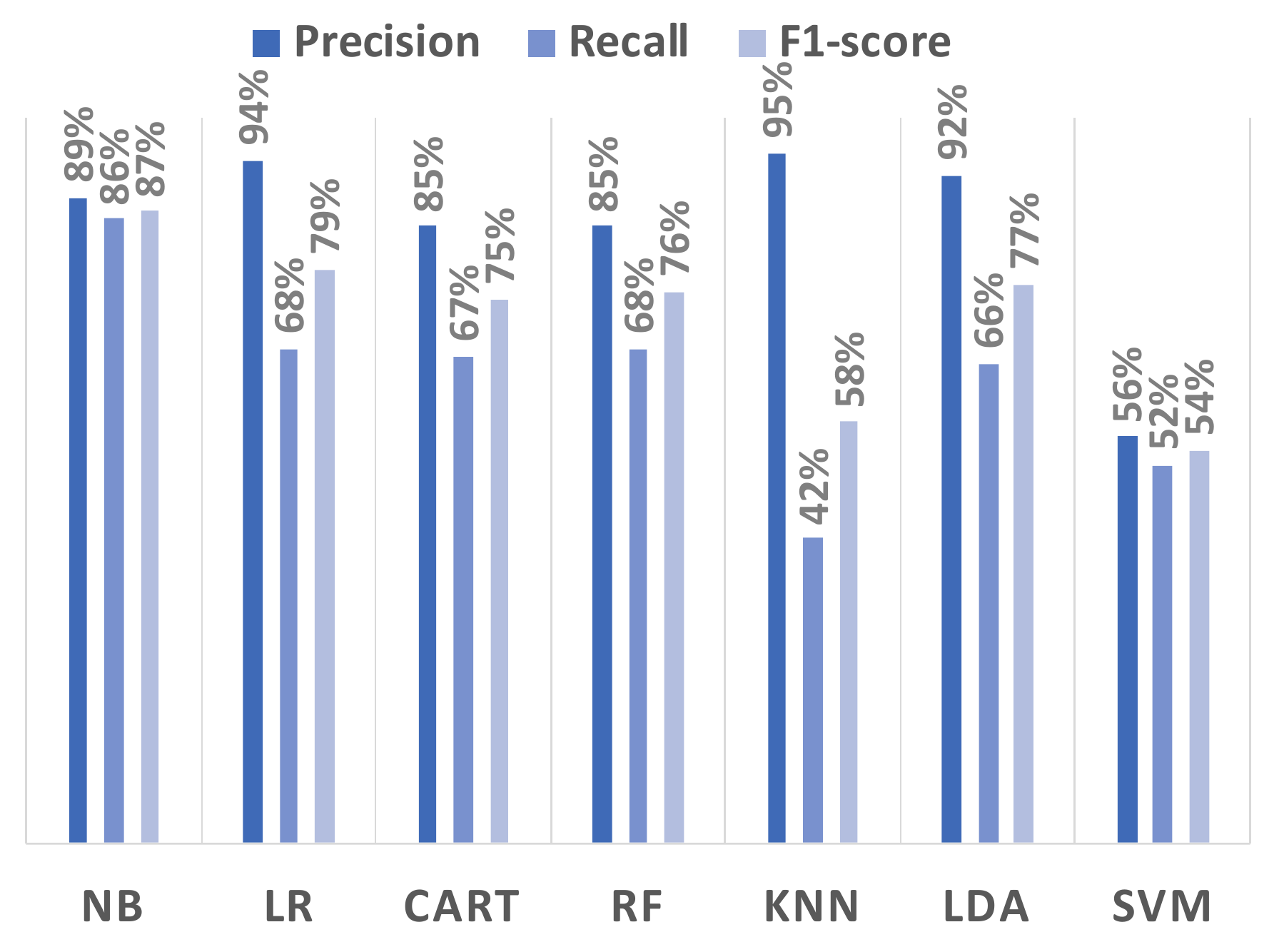}
    \caption{Naive Bayes tops by performance (Accuracy, Precision, Recall and F1-score) comparison among NB, LR, CART, RF, KNN, LDA and SVM on dataset LD137.}
    \label{fig:model_comparison}
\end{figure}

\textbf{3. Fixing the number of words per field.}
We want to identify the most discerning words from each text field, which is a standard process in NLP for  improving the scalability, efficiency and accuracy of a text classifier \cite{chen2009feature}.
Using the ${\chi}^2$ statistic, we select the top $K_f$ best words from each field. 

To select the appropriate number of words per field,
we followed the process below.
We vary $K_f$ = 5,10,20,30,40 and 50 for title, topic and README file, and 
we find that the top 30 words in title, 10 words in topic and 10 words in README file exhibit the highest accuracy. Similarly, we try $K_f$ = 80, 90, 100, 110 and 120 for file names and $K_f$ = 300, 325, 350, 375, 400, 425, 450 and 475 for the description field. We find that the top 100 words for file and folder names and top 400 words for description field give the highest accuracy. 
Note that we do this during training and refining the algorithm,
and then we continue to use these words as features in testing.

Thus, we select the top: (a) 30 words from the title, (b) 10 words from the topics, (c) 400 words from the description, (d) 100 words from the file names, and 10 words from the README file. 
This leads to a total of  550 words across all fields.
For reference, we find 9253 unique words  in the repository fields of our training dataset.
Reducing the focus on the top 550 most discerning words per field
 increases the classification accuracy by as much as 20\% in some cases.


{\bf 4. Evaluating and selecting ML algorithms:}
 We asses the classification performance of Multinomial Naive Bayes (NB), Logistic Regression (LR), Support Vector Machine (SVM), Decision Tree (CART), Random Forest(RF),  Linear Discriminant Analysis (LDA), and K-Nearest-Neighbors (KNN),
 and show  their precision, recall and F1-score in Figure~\ref{fig:model_comparison}. 
 
 Multinomial Naive Bayes exhibits the best  F1-score with \fOne,
 striking a good balance between \precision precision, \recall recall for the malware class.
 Detecting the benign class we do even better with 92\% precision, 94\% recall and 93\% F1-score.
 By contrast, the F1-score of the other algorithms is below 79\%.
 Note that KNN, LR and LDA provide higher precision, but with significantly lower recall. Clearly, one could use these algorithms 
 to get higher precision at the cost of lower total number of repositories. 

We use Multinomial Naive Bayes as our classification engine for the rest of this study. 
We attempt to explain the superior F1-Score of the Naive Bayes in our context. The main advantage of Naive Bayes over other algorithms is that it considers the features independently of each other and can handle large number of features better. As a result, it is more robust to noisy or unreliable features.
It  also performs well in domains with many equally important features, where other approaches suffer, especially with a small training data, and it is not prone to overfitting~\cite{ting2011naive}. As a result, the Naive Bayes is considered a dependable algorithm for text classification and it is often used as the benchmark to beat~\cite{xu2018bayesian}.

\begin{figure}
    
    
    \includegraphics[width=0.99\linewidth, height=5cm]{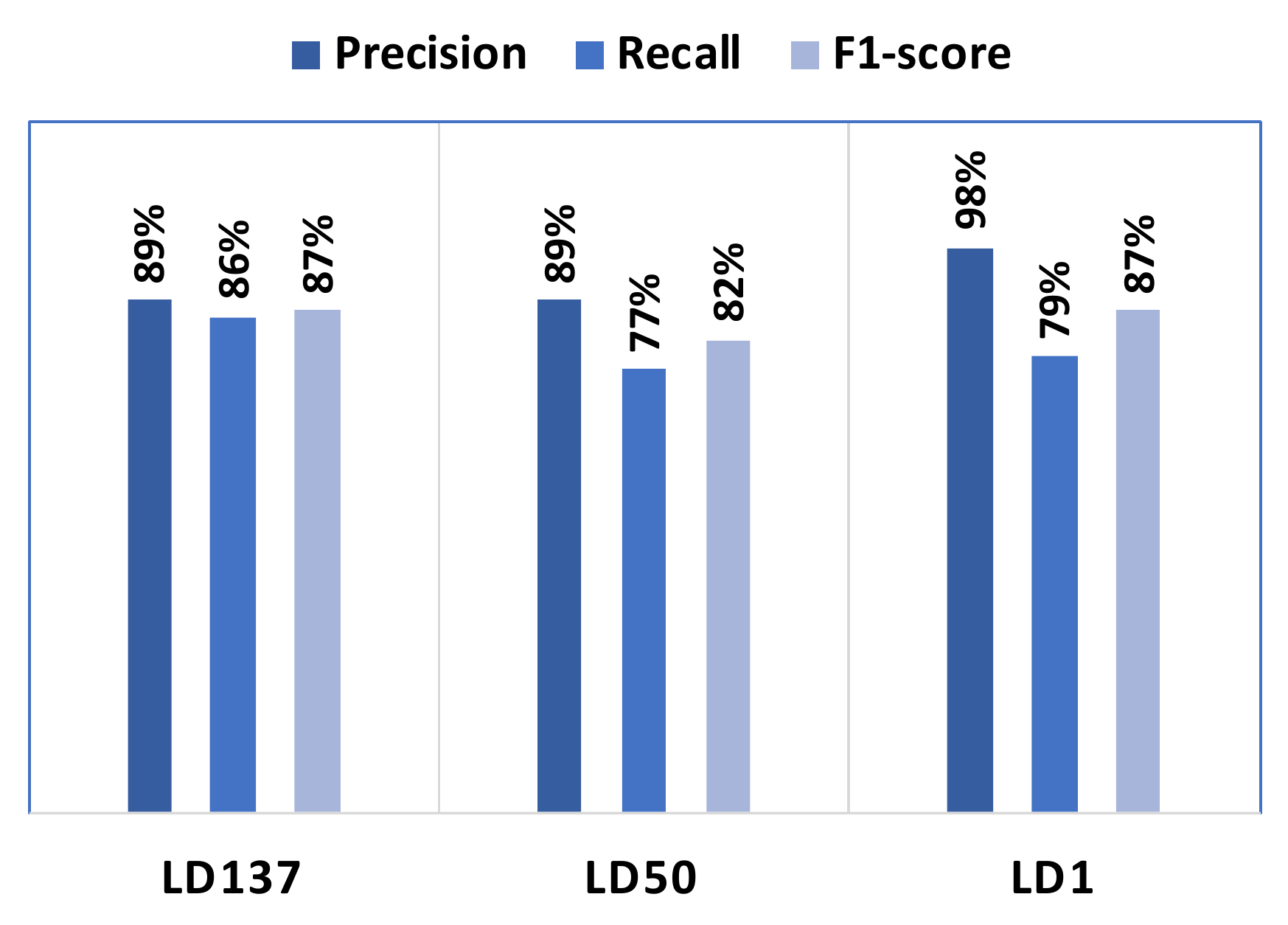}
    
    
    \caption{Assessing the effect of the number of keywords in the query: Precision, Recall and F1-score of our approach on the LD137, LD50 and LD1 labeled datasets.}
    \label{fig:precision-recall}
\end{figure}

{\bf 5. Assessing the effect of the query set:}
We have made the following choices in the previous steps:
(a) 5 text-based fields,
(b) bag of words with count vectorization,
(c) 550 total words across all the fields,
and (d) the Multinomial Naive Bayes.
We perform 10-fold cross validation and report the precision, recall and F1-score in Figure~\ref{fig:precision-recall} for our three different labeled data sets. 
We see that the 
 precision stays above 89\% for all three datasets,
 with a recall above 77\%.

It is worth noting the relative stability of our approach with respect to the keyword set for the initial query especially between LD50 and LD137 datasets.
The LD1 dataset we observe higher accuracy, but significantly less recall compared to LD137. We attribute this fact to the single keyword used in selecting the repositories in LD1, which may have lead to a more homogeneous group of repositories. 
Interestingly, LD50 seem to have the lower recall and F1-score
even though the differences are not that large.


\textbf{6. Identifying  \malrepo malware repositories:}
We use LD137 to train our Multinomial Naive Bayes model and apply it on RD137 dataset. We find \malrepo malware repositories.
We also apply the same trained model on  RD1 and RD50 and find 809 and 3615 malware repositories respectively, but this repositories are included in the \malrepo. (Recall that RD1 and RD50 are subsets of RD137). 

\begin{table}
    \centering
    \footnotesize
    \begin{tabular}
        {|p{0.14\linewidth}|p{0.13\linewidth}||p{0.18\linewidth}|p{0.32\linewidth}|}
        \hline
         \textbf{Dataset} & \textbf{Initial} & \textbf{Malware} & \textbf{Mal. + Source} \\ \hline
         RD1 & 2775 & 809 & 680 \\ \hline
         RD50 & 14332 & 3615 & 3096 \\ \hline
         RD137 & 97375 & 8644 & 7504  \\ \hline
    \end{tabular}
    \caption{The identified  repositories  per dataset with: (a) malware, and (b) malware and source code.}
    \label{tab:mal-identifcation}
\end{table}

\textbf{7. Identifying  \sourceNum malware source code repositories:}
We use our heuristic approach to  identify  source code repositories.
We set our \sourceThresh threshold to 75\%, meaning that:
if more than 75\% of files in a repository are source code files, we label it as a source code repository. Applying this heuristic, we find that \sourceNum repositories
are most likely  source code repositories in RD137. We use the name \textbf{M137} to refer to this group of  malware source code repositories. We find 680 and 3096 malware source code repositories in RD1 and RD50 as shown in Table \ref{tab:mal-identifcation}. However, these are subset of M137, given that RD1 and RD50 are subsets of RD137. 

To evaluate the effectiveness of our heuristic,  we manually check  30 randomly-selected repositories from M137. We find that all 30 repositories contain source code\footnote{Apart from a manual verification, these 30 repositories were further stress-tested: (a) 20 where used in a separate static analysis study, and (b) 15 were compiled and run successfully within an emulator.
},
which corresponds to 100\% precision.
We will further evaluate the effectiveness of this heuristic in the future.






\textbf{8. A curated malware source code dataset: \curDataSet}
As a tangible contribution,
we provide, \curDataSet, a  dataset of 250  repositories from the M137 dataset, which we manually verify  for containing malware source code and
relating to a particular  malware type.
Opting for diversity and coverage,
the dataset spans all the identified types: virus, backdoor, botnet, keylogger, worm, ransomware, rootkit, trojan, spyware, spoof, ddos, sniff, spam, and cryptominer. 
While  constantly updating, we will make this dataset available to researchers.




\section{A large scale study of malware}
\label{sec:longitudinal}
\vspace{-0.15in}
Encouraged by the substantial number of malware repositories,
we study the distributions and longitudinal properties of the 
identified malware repositories  in M137. 

{\bf Caveat:} We provide some key observations in this section, 
but they should be viewed as indicative and approximate trends
and only within the context of the collected repositories
and with the general assumption that repository titles and
descriptions are reasonably accurate.
In Section~\ref{sec:discussion}, we discuss issues around the 
biases and limitations that our dataset may introduce.

    
\begin{figure}[h]
    \centering
    \includegraphics[height=5cm,width=0.40\textwidth]{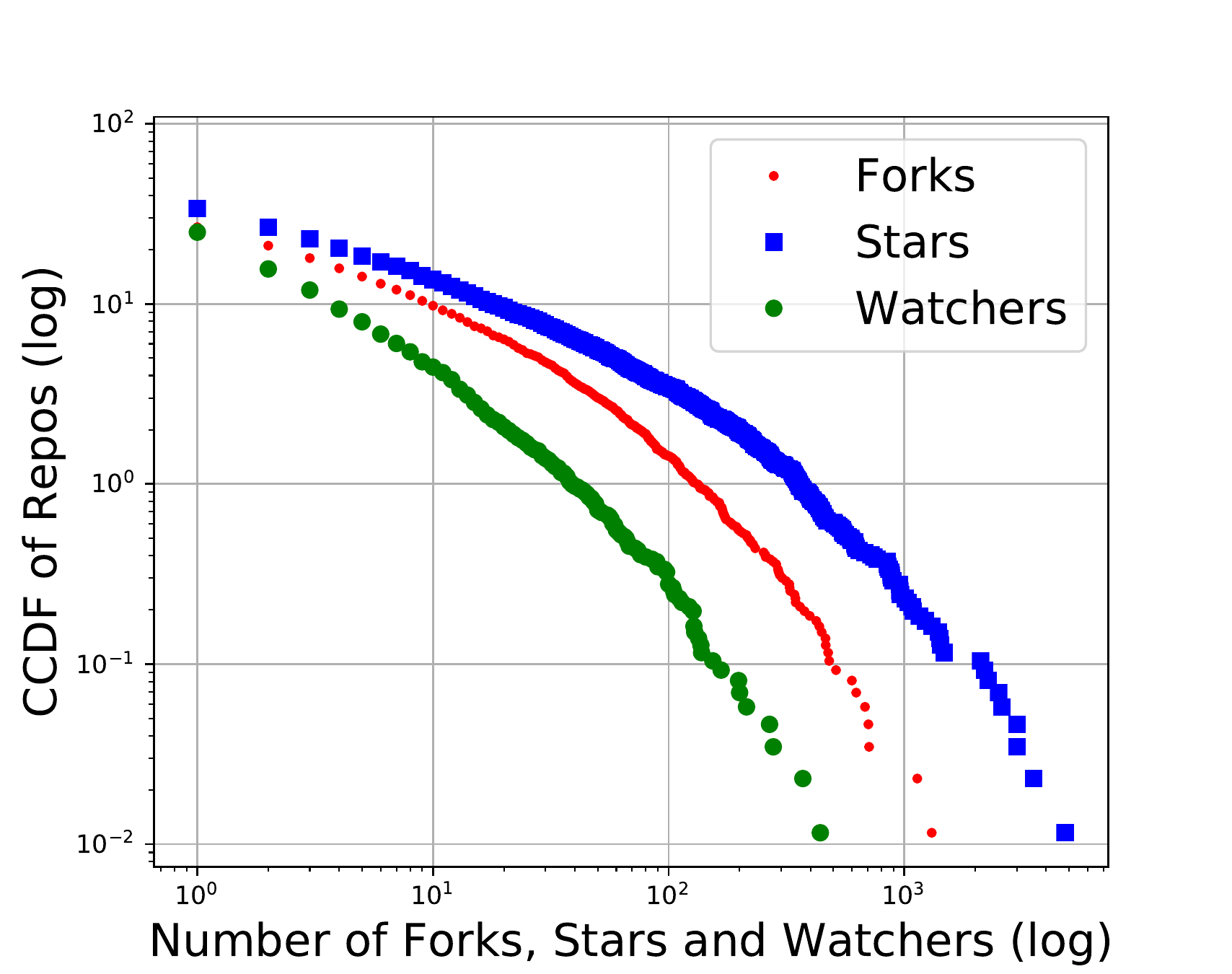}
    \caption{CCDF distributions of forks, stars and watchers per repository.}
     \label{fig:ccdf-repos}
\end{figure}

\textbf{A. Identifying influential repositories.}
The prominence of a repository can be measured by the number of \textit{forks}, 
\textit{stars}, and \textit{watchers}.
In Figure~\ref{fig:ccdf-repos}, we plot the complementary 
cumulative distribution function (CCDF) of these three metrics for our
malware repositories.


{\bf Fork distribution:}  We find that 2\% of the repositories 
seem quite influential with at least 100 forks
as shown in  Figure~\ref{fig:ccdf-repos}.
Recall that the fork counter indicates the number of distinct users that have forked a repository.
For reference, 78\% of the repositories have less
than 2 forks. 


{\bf Star distribution}: We find that 2\% of the 
repositories receive more than 250 stars as shown in Figure~\ref{fig:ccdf-repos}. For reference, 
75\% of the repositories have less than 3 stars.

{\bf Watcher distribution}: In
Figure~\ref{fig:ccdf-repos}, we find that 1\% of the 
repositories have more than 50 watchers. For reference, 
we  observe that 84\% of
the repositories have less than 3 watchers.
Note that these distributions are skewed, and follow patterns
that can be approximated by a log-normal distribution. 

{\em Which are the most influential repositories?}
We find that 8 repositories dominate the top 5 spots across all three metrics:  stars, forks, and  watchers. We present a short profile of these dominant repositories in Table \ref{tab:top8repo}. Most of the repositories contain a single malware project, which is an established practice among the authors in \github~\cite{projectperrepo1,projectperrepo2}. We find that the repository ``theZoo''~\cite{thezoo}, created by \textit{ytisf} in 2014
is the most forked, watched, and starred repository with 1393 forks, 730 watchers and 4851 stars as of October, 2019. 
However, this repository is quite unique and was created with the intention of being a malware database with 140 binaries and 80 source code repositories.

\begin{table}[t]
    \centering
    \footnotesize
    \begin{tabular}
        {|p{0.02\linewidth}|p{0.16\linewidth}|p{0.05\linewidth}|p{0.05\linewidth}|p{0.1\linewidth}|p{0.32\linewidth}|}
        \hline
         \textbf{R ID} & \textbf{Author} & \textbf{\# Star} & \textbf{\# Fork}& \textbf{\# Watcher}& \textbf{ Content of the Repository}\\ \hline
         1 & ytisf & 4851 & 1393 & 730 &  80 malware source code and 140 Binaries\\ \hline
         2 & n1nj4sec & 4811 & 1307 & 440 & Pupy RAT\\ \hline
         3 & Screetsec & 3010 & 1135 & 380 &  TheFatRat Backdoor\\ \hline
         4 & malwaredllc & 2515 & 513 & 268 &  Byob botnet\\ \hline
         5 & RoganDawes & 2515 & 513 & 268 &  USB attack platform\\ \hline
         6 & Visgean & 626 & 599 &  127 &  Zeus trojan horse\\ \hline
         7 & Ramadhan & 535 & 283 & 22 &  30 malware samples\\ \hline
         8 & dana-at-cp & 1320 & 513 & 125 &  backdoor-apk backdoor\\ \hline

    \end{tabular}
    \caption{The profile of the top 5 most influential malware repositories across all three metrics with 8 unique repositories.}
    \label{tab:top8repo}
\end{table}



{\bf Influence metrics are correlated}: 
As one would expect, the influence and popularity metrics are correlated.
We use a common correlation metric, 
the Pearson Correlation Coefficient ($r$)~\cite{benesty2009pearson}, measured in a scale of $[-1,1]$. We calculate the metric for all pairs of our three popularity metrics. We find that all of them exhibit  higher positive correlation: stars vs. forks ($r=0.92$, $p<0.01$), forks vs. watchers ($r=0.91$, $p<0.01$) and watchers vs. stars ($r=0.91$, $p<0.01$).


\textbf{B. Malware Type and Target Platform.}
We wanted to get a feel for what type of malware we have identified.
As a first approximation,
we use the keywords found in the text fields  to relate repositories  in M137 with the type of malware and the 
intended target platform.
Our goal is to create
 the two-dimensional distribution per
 malware type and the target platform as shown in
 Table~\ref{tab:type_os_distribution}.
To create this
table, 
we associate a repository
with keywords in its title, topics, descriptions, file names and README file fields of: (a) the 6 target platforms,
and (b) the 13 malware type keywords.

\textit{How well does this heuristic approach work?}
We provide two different indications of its relative effectiveness.
First, the vast majority of the repositories relate to
one platform or type of malware:
(a) less than 8\% relate to more than one platform,
and (b) less than 11\% relate to more than one type of malware.
Second, we manually verify the  250 
repositories in our curated data~\curDataSet
and find a 98\% accuracy.



Below, we provide some observations from 
Table~\ref{tab:type_os_distribution}.

{\bf a. Keyloggers reign supreme.} We see that one of the largest 
categories is the keylogger malware with 679 repositories, which 
are mostly affiliated with Windows and Linux platforms.
We discuss the emergence of keyloggers
below in  our temporal analysis.

{\bf b. Windows and Linux are the most popular targets.}
Not surprisingly, we find that the majority of the malware repositories  
are affiliated with these two platforms: 1592 repositories for 
Windows, and 1365 for Linux.  

\begin{figure*}[h]
    \begin{subfigure}{0.26\textwidth}
    \includegraphics[height=5cm,width=5cm]{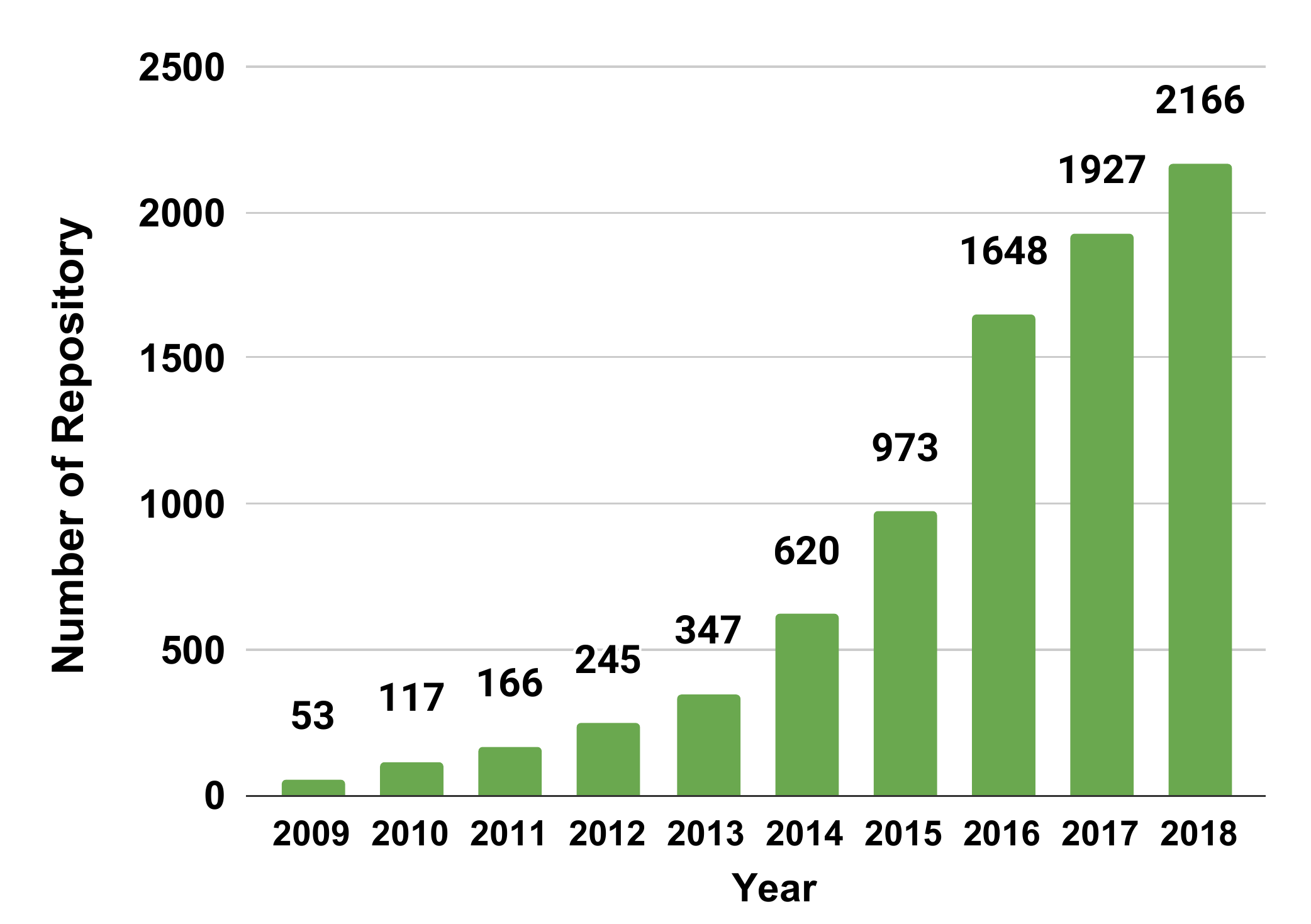}
    
    \caption{New malware repositories created per year.}
    \label{fig:overall-trend}
    \end{subfigure}
    ~
    \begin{subfigure}{0.42\textwidth}
    \includegraphics[height=5cm,width=7.7cm]{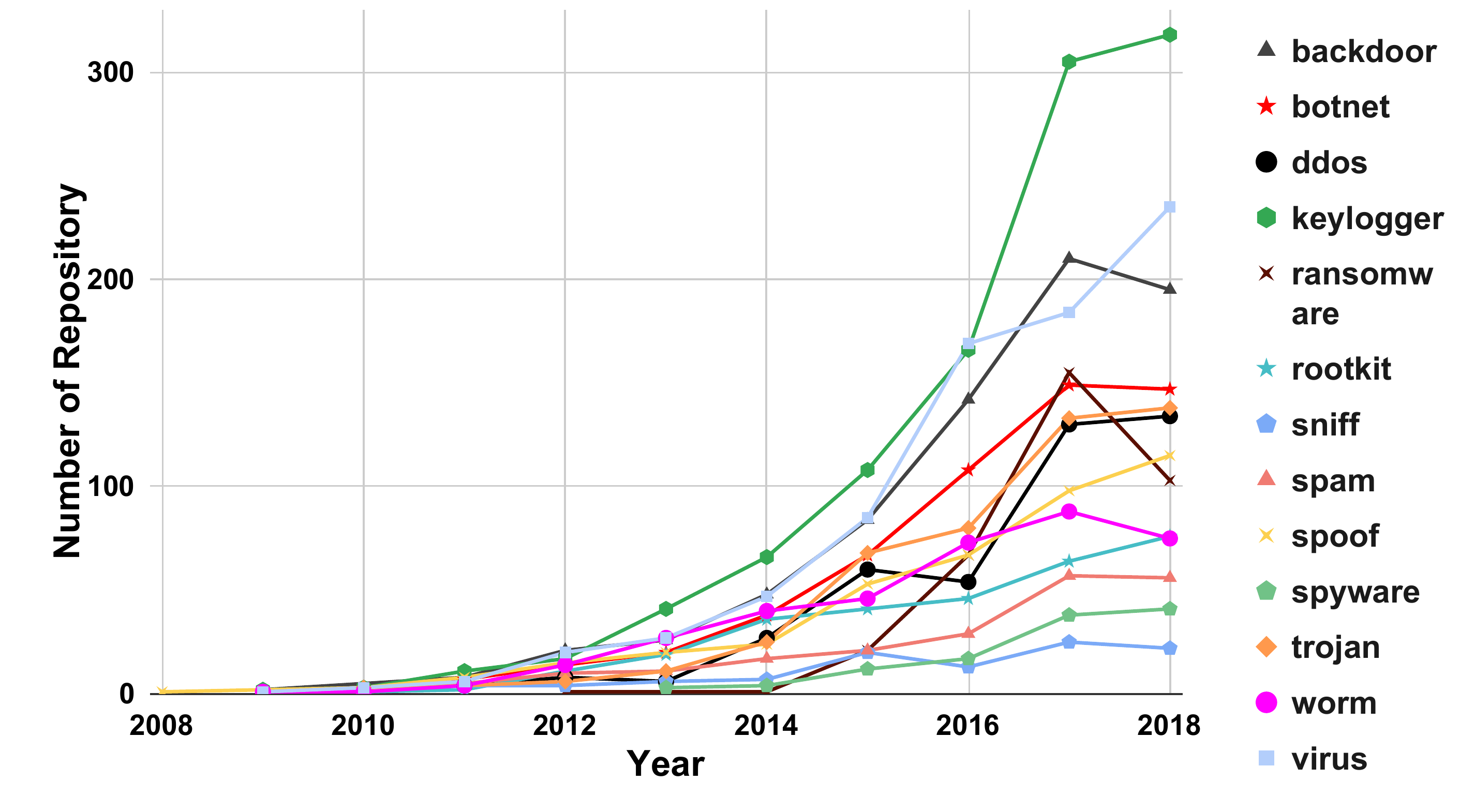}
    
    \caption{New repositories per type of malware per year.}
    \label{fig:type-trend}
    \end{subfigure}
    ~
    \begin{subfigure}{0.30\textwidth}
    \includegraphics[height=5cm,width=5.5cm]{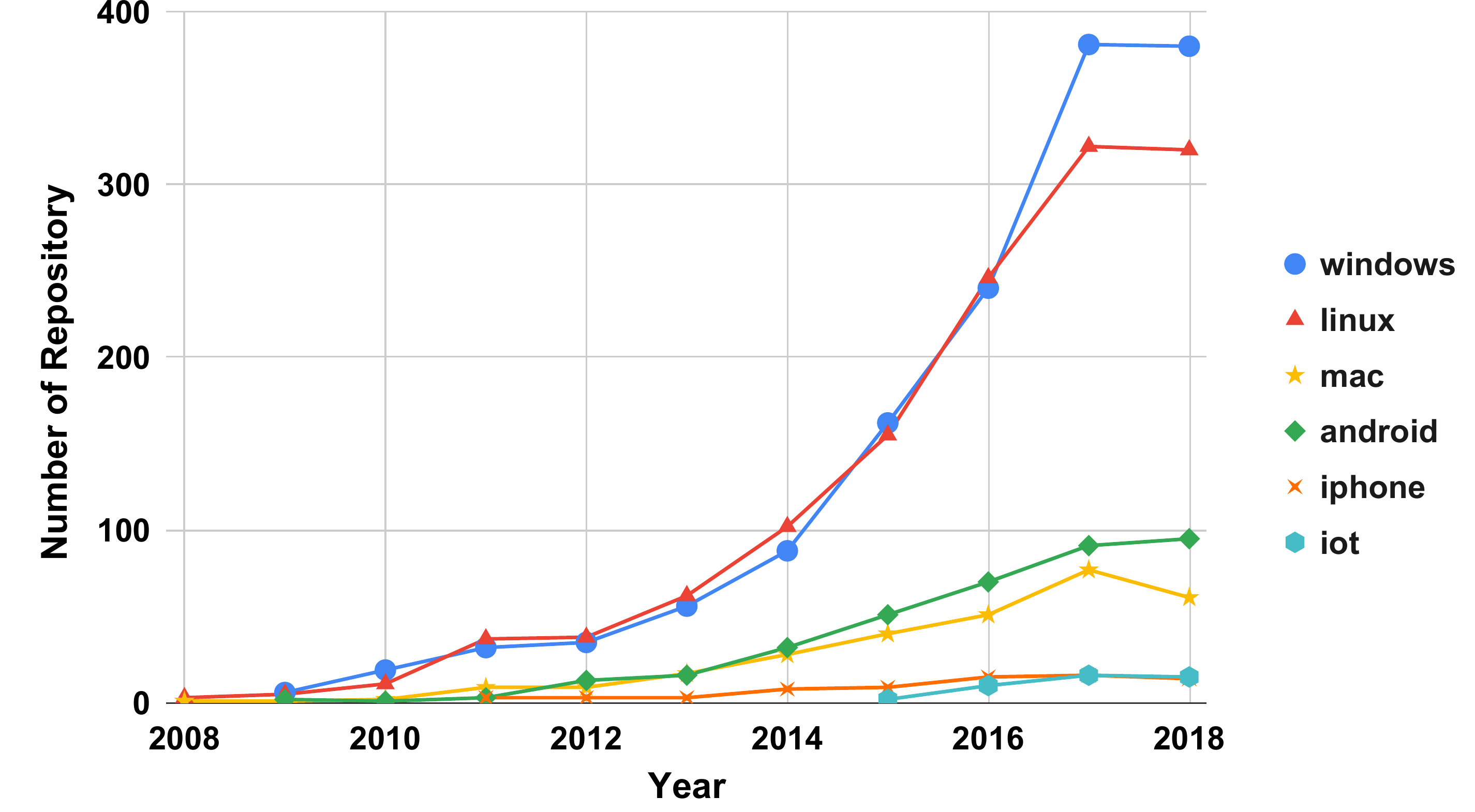}
    
    \caption{New  malware repositories per target platform per year.}
    \label{fig:platform-trend}
    \end{subfigure}
    
    \caption{New  malware repositories per year: a) all malware,  b) per type of malware, and c)  per target platform. 
    }
    \label{fig:trend}
\end{figure*}

{\bf c. MacOS-focused repositories: fewer, but they exist.} 
Although MacOS platform are less common among PC users,
we see that malware repositories targeting such platforms indeed
exist. As shown in Figure~\ref{fig:platform-trend}, 
MacOS malware repositories are an order of magnitude less compared
to those for Windows and Linux.
\begin{table}[t]
    \footnotesize
    \begin{tabular}
        {|p{0.10\linewidth}|r|r|r|r|r|r|r|}
        \hline
        \multirow{2}{*}{\textbf{Types}} & \multicolumn{7}{|c|}{\textbf{Target Platform}} \\
        \cline{2-8}
        
         & Wind. & Linux & Mac & IoT & Andr. & iOS & {\bf Total} \\
        \hline \hline
        {\bf Total} & {\bf 1592} & {\bf 1365} & {\bf 380} & {\bf 108} & {\bf 442} & {\bf 131} & {\bf 4018} \\ \hline \hline
        key-
        logger & 396 & 209 & 42 & 2 & 27 & 3 & {\bf 679} \\ \hline
         back-
        door & 181 & 227 & 37 & 11 & 51 & 4 & {\bf 511} \\ \hline
        virus & 235 & 131 & 34 & 2 & 51 & 16 & {\bf 469} \\ \hline
       
        botnet & 153 & 154 & 43 & 36 & 64 & 17 & {\bf 467} \\ \hline
        trojan & 133 & 70 & 24 & 16 & 67 & 19 & {\bf 329} \\ \hline
        spoof & 76 & 115 & 88 & 2 & 20 & 9 & {\bf 310} \\ \hline
        rootkit & 55 & 163 & 13 & 2 & 19 & 3 & {\bf 255} \\ \hline
        
        ransom-
        ware & 117 & 67 & 14 & 1 & 33 & 13 & {\bf 245} \\ \hline
        ddos & 71 & 95 & 20 & 10 & 9 & 3 & {\bf 208} \\ \hline
        worm & 61 & 45 & 18 & 5 & 25 & 18 & {\bf 172} \\ \hline
        spyware & 45 & 22 & 6 & 6 & 38 & 16 & {\bf 133} \\ \hline
        
        spam & 40 & 29 & 18 & 14 & 23 & 5 & {\bf 129}  \\ \hline
        sniff & 29 & 38 & 23 & 1 & 15 & 5 & {\bf 111} \\ \hline

    \end{tabular}
    \caption{Distribution of the malware repositories from M137
    dataset based on the malware type and malware target platform.
    This table demonstrates the repositories that fit with the
    criteria defined in Section~\ref{sec:longitudinal}.}
    \label{tab:type_os_distribution}
\end{table}

\textbf{C. Temporal analysis.}
We  want to study the evolution 
and the trends of malware repositories. 
We plot the number of new malware repositories per year:  a) total malware, b) per type of malware, and c) per target platform in Figure~\ref{fig:trend}.
We discuss a few interesting temporal behaviors below. 


{\bf a. The number of new malware repositories more than triples every four years.}
We see an alarming increase from 117 malware repositories in 2010 to 620 repositories in 2014 
and to 2166 repositories in 2018.
We also observe a sharp increase of 70\% between 2015 to 2016 shown in Figure~\ref{fig:overall-trend}.


{\bf b. Keyloggers  started a super-linear growth since 2010} 
and are by far affiliated with the most new repositories per year since 2013, but their rate of growth reduced in 2018.

 {\bf c. Ransomware repositories emerge in 2014 and 
 gain momentum in 2017}. 
 Ransomware experienced their highest growth rate in 2017 with 155 new repositories, while that number dropped to 103 in 2018.
 

{\bf d. Malware activity slowed down in 2018 across the board.}
 It seems that 2018 is a slower year for all  malware 
 even when seen by type
( Figure~\ref{fig:type-trend})
 and target platform (Figure~\ref{fig:platform-trend}).
We find that the number of new malware repositories has dropped significantly in 2018 for most types of malware except virus, keylogger and trojan.

{\bf e. IoT  and iPhone malware repositories become more visible after 2014}.
We find that IoT malware emerges in 2015 and iPhone malware sees an increase after 2014 in Figure~\ref{fig:platform-trend}. 
We conjecture that
this is possibly encouraged by the emergence and increasing popularity of 
specific malware: (a) WireLurker, Masque, AppBuyer malware~\cite{iosmalware2014} for iPhones, and (b) BASHLITE \cite{iotbashlite}, a Linux based botnet for IoT devices.
We find the names of  the aferemntioned malware  in many repositories starting in 2014.
Interestingly, the source code of the original BASHLITE botnet is available in a repository created by {\em anthonygtellez} in 2015.



{\bf f. Windows and Linux: dominant but slowing down.} In Figure \ref{fig:platform-trend}, we see that  windows and linux malware are flattened between 2017 and 2018. By contrast, IoT and android repositories have increased.






\section{Understanding malware authors}
\label{sec:authors}
\vspace{-0.15in}
Intrigued by the fact that authors create public malware repositories, we attempt to understand and profile their behavior.

As a first step towards understanding the \malauthor, we
want to assess their popularity and influence. 
We  use  the following  metrics:
(a) number of malware repositories which they created,
(b) number of followers,
(c) total number of watchers on their repositories,
and (d) total number of stars.
We  focus on the first two metrics here.
We  use the notation 
  \textit{top k authors} for any of the metrics above, where \textit{k} can be any positive integer to referring to "heavy-hitters". 


\textbf{A. Finding influential \malauthor. } 
We study the distribution of the number of malware repositories created and the number followers per author in following.

\begin{figure}[]
    \centering
    \includegraphics[height=4cm, width=0.40\textwidth]{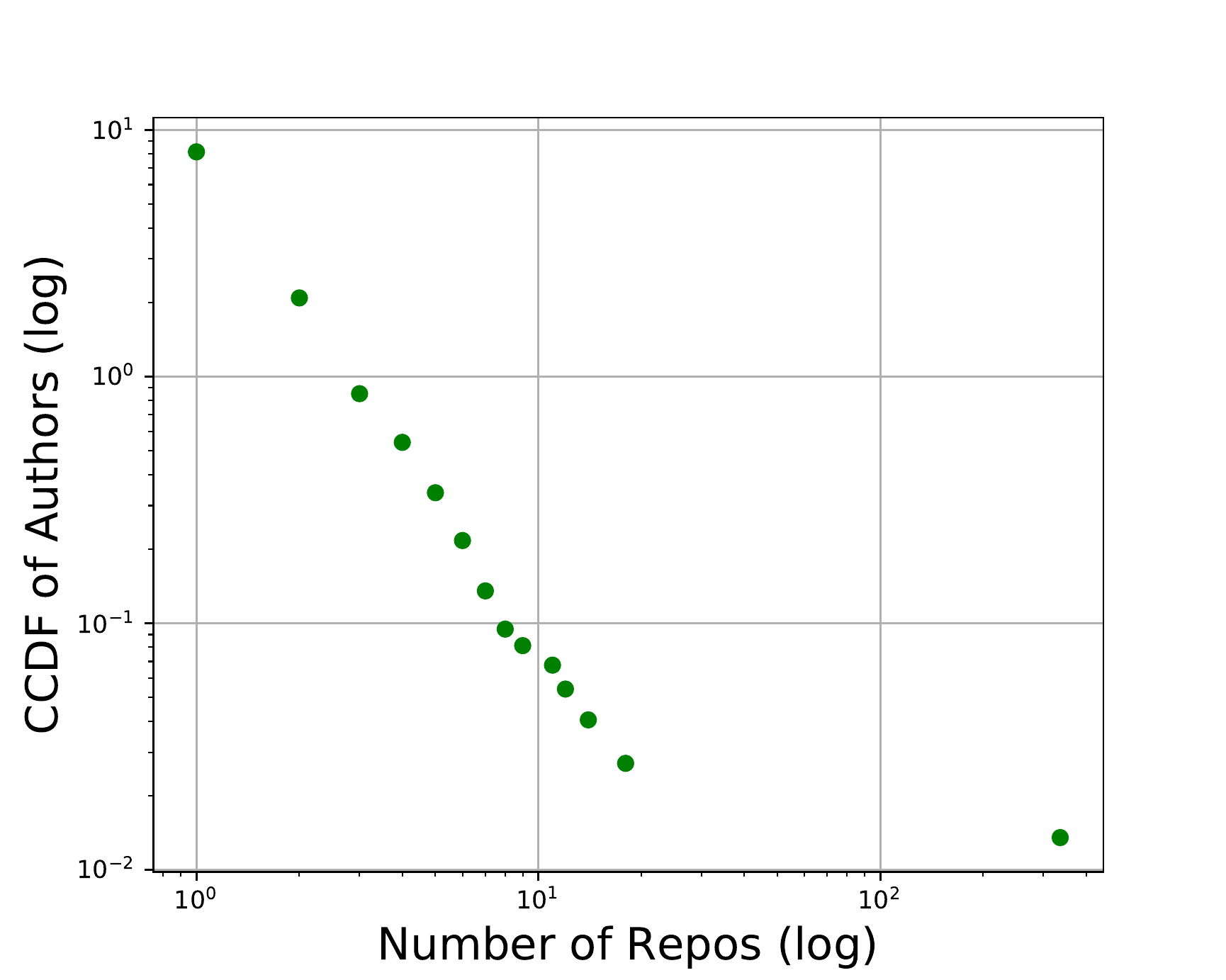}
    \caption{CCDF of malware repositories per author.}     \label{fig:cddf-author}
\end{figure}

First, we find that 
15 authors are contributing roughly 5\% of all malware repositories
by examining the CCDF of the created repositories in Figure~\ref{fig:cddf-author}. 
From the figure, we find an outlier author, {\em cyberthreats}, who doesn't follow power law distribution~\cite{faloutsos1999power}, 
has created 336 malware repositories.
We also find that 99\% authors have less than 5 repositories. 


Second, we study the distribution of the number of followers per author but omit the plot due to space limitations. The distributions is skewed with
  3\% (221) of the authors  having more than 300 followers each,
  while  70\% of the authors have less than 16 followers.
    
\textbf{B. \MalAuthors strive for an online ``brand":}
In an effort to understand the motive of sharing
malware repositories, we make the following investigation. 

{\bf a. Usernames seem persistent across online platforms.}
We find that many \malauthor 
use the same username consistently across many
 online platforms, such as security forums.
We conjecture that they are developing a reputation
and they use their username as a ``unique" identifier.

 We identify \malHackersNo \malauthor\footnote
 {
 Note that this does not mean that the other authors are not doing the same, but they could be active in other security forums or online platforms.
 },
who are active in at least one of the three 
security forums: Offensive Community, Ethical Hacker and Hack This Site, for which we happen to have access to their data. We conjecture that at least some of these usernames correspond to the same users based on the following two indications.
First, we find direct connections between the usernames across different platforms. For example,
user {\em 3vilp4wn} at the
 ``Hack This Site'' forum  is promoting a keylogger malware  
 by referring to a
\github repository~\cite{hacktool3vilp4wn}
whose author has the same username. 
Second, these usernames are fairly uncommon, which 
increases the likelihood of belonging to the same person.
For example,
 there is a \github user with the name {\em fahimmagsi},
 and someone with the same username is boasting about their hacking successes in the ``Ethical Hacker'' forum.
 As we will see below, {\em fahimmagsi} seems to
 have a well-established online reputation.
 
{\bf b. ``Googling" usernames reveals significant hacking activities.} 
Given that these \github usernames are fairly unique, it was natural
to look them up on the web at large.
Even a simple Internet
search with the  usernames reveals 
significant hacking activities, including hacking 
websites or social networks, and offering hacking tutorials in YouTube. 

We  
 investigate  the \textit{top 40} most prolific \malauthor using a web search with a {\em single} simple query:  ``hacked by $<$username$>$''.
We then examine only the first page of search results.
Despite all these self-imposed restrictions,
we identify three users with substantial hacking related activities across Internet.
For example,
we find  a number of news articles for hacking a series of websites by  \github users {\em fahimmagsi} and {\em CR4SH} \cite{hacknewsfahimmagsi}~\cite{hacktoolcr4sh}. Moreover, we find user {\em n1nj4sec} sharing a multi-functional Remote Access Trojan (RAT) named ``Pupy", developed by her, which received significant news coverage in security articles back in March of 2019~\cite{pupyGithub}\cite{pupyarticle}.
We are confident that well-crafted and targeted searches can connect
more malware authors with hacking activities and usernames in other online forums.

\vspace{-0.15in}
\section{Discussion} 
\label{sec:discussion}
\vspace{-0.15in}
We discuss the effectiveness and limitations of \algo.

\textbf{a. Why is malware publicly available in the first place?}
Our investigation in Section~\ref{sec:authors} provides
strong indications that malware
 authors want to actively establish their hacking reputation.
 It seems that they want  to boost their online credibility, which often translates to money. Recent works~\cite{Portnoff2017, Deb2018_USC3,Sapienza2018_USC2} study the
 underground markets of malware services and tools:
 it stands to reason that notorious hackers will attract more clients. 
At the same time, \github acts as a collaboration platform, which can help hackers improve their  tools. 

\textbf{b. Do we identify every malware repository in \github?} Our tool can not guarantee that it will identify every malware repository in \github. First, 
we can only identify repositories that ``want to be found":
(a) they must be public, and (b) they must be described with the
appropriate text and keywords.
Clearly, if the author wants to hide her repository,  we won't be able to find it.
However, we argue that this defeats the purpose of having a public archive: if secrecy was desired, the code would have been shared through private links and services.
Second, our approach is constrained by \github querying limitations,
which we discussed in Section \ref{sub-sec:data-collection}, and the set of 137 keywords that we use. 
However, we are encouraged by the number and the reasonable diversity
of the retrieved repositories we see in Table~\ref{tab:type_os_distribution}. 

\textbf{c. Are our datasets representative?} This is the typical hard question for any measurement or data collection study. 
First of all, we want to clarify that our goal 
is to create a large database of malware source code.
So, in that regard, we claim that we accomplished our mission.
At the same time, we seem
to have a fair number of malware samples in each category of interest, as we see in Table~\ref{tab:type_os_distribution}.

Studying the trends of malware is a distant second goal, which we present with the appropriate caveat.
On the one hand, we are limited by \github's API operation, as we discussed earlier.
On the other hand, we attempt to reduce the biases that are
under our control.
To ensure some diversity among our malware,
 we added as many words as we could in our 137 malware, which is likely to capture a wide range of malware types. 
We argue that the fairly wide breadth of malware types  in Table \ref{tab:type_os_distribution} is a good indication.
Note that our curated dataset \curDataSet with 250 malware is reasonably representative in terms of coverage. 

{\bf d. What is the overlap among the identified repositories?}
Note that our repository does not include forked repositories, 
since {\bf \github does not return  forked repositories as answers to a query}. 
Similarly, the breadth of the types of the malware as shown in Table~\ref{tab:type_os_distribution} hints at a reasonable diversity.
However, our tool cannot claim that the identified repositories are distinct nor is it attempting do so.
\github does not restrict  authors from copying (downloading),
 and uploading it as a new repository. 
In the future, we intend to study the similarity and evolution among these repositories.

{\bf e. Are the authors of repositories the original creator of the source code?}
This is an interesting and complex question that goes beyond the scope of this work.
 Identifying the original creator will require studying the source code of all related repositories, and analyzing the dynamics of the hacker authors, which we intend to do in the future. 

 {\bf f. Are all the malware authors malicious?}
Not necessarily. This is an interesting question, but it is not
central to the main point of our work.
On the one hand, 
we find  some white hackers or researchers, such as Yuval Nativ~\cite{yuval}, or Nicolas Verdier~\cite{Nicolas}.
On the other hand, several authors seem to be malicious, as we  saw in Section~\ref{sec:authors}. 

{\bf g. Are our malware repositories in "working order"?} 
It is hard to know for sure, but we attempt to answer indirectly.
First, we pick 30 malware source codes and all of them compiled
and a subset of 15 of them actually run successfully in an emulated environment as we already mentioned.
Second, these public repositories are a showcase for the skills 
of the author,
who will be  reluctant to have repositories of low quality. 
Third, public repositories, especially popular ones, 
are inevitably scrutinized by their followers.

\textbf{h. Can we handle evasion efforts?} Our goal 
is to create the largest malware source-code database possible
and having collected \sourceNum malware repositories seems like a great start.
In the future, 
\malauthor could obfuscate their repositories by using misleading titles, and description, and even filenames.
We argue that authors seem to want their repositories to be found,  which is why they are public. 
We also have to be clear: it is easy for the authors to  hide their repositories, and they could would start by making them private or avoid \github altogether. 
However, both these moves will diminish the visibility of the authors.


\textbf{i. Will our approach generalize to other archives? } 
We believe that \algo can generalize to other archives, which 
provide public repositories, like GitLab and BitBucket. We find that these
sites allow public repositories and let the users retrieve
repositories. We have also seen equivalent 
data fields (title, description, etc). Therefore, we are 
confident
that our approach can work with other archives.



\section{Related Work}
\label{sec:related}
\vspace{-0.15in}
There are several works that attempt to determine if a piece of software is malware, usually focusing on a binary,  using static or dynamic analysis~\cite{aycock2006computer,kolbitsch2009effective,shankarapani2011malware,damodaran2017comparison}. However,
to the best of our knowledge, no previous study has focused on 
identifying malware source code in public software archives, such 
as \github, in a systematic manner as we do in this work.
We highlight the related works in the following categories:

{\bf a. Studies that needed source code.}
Several 
studies~\cite{lepik2018art,zhong2015stealthy,shen2018javascript} 
use malware source code that are manually retrieved from 
\github repositories. 
Some studies~\cite{calleja2016look}~\cite{calleja2018malsource} compare the evolution and the code reuse of 150  malware source codes (with only some from \github) with that of benign software from a software engineering perspective and  study the code reuse.  
Overall, various studies  
\cite{darki2019idapro,jerkins2017motivating} can benefit from 
malware source code to fine-tune their approach. 

\textbf{b. Mining and analyzing \github:}
Many studies have analyzed different aspects of \github, but not with
the intention of retrieving malware repositories.
First, there are efforts that 
study the user interactions and collaborations on \github and their 
relationship to other social media in
\cite{kollanyi2016automation,horawalavithana2019mentions,pletea2014security}.
Second, some efforts discuss the challenges in extracting 
and analyzing data from \github with respect to sampling 
biases~\cite{cosentino2016findings,gousios2017mining}. 
Other works~\cite{kalliamvakou2014promises,kalliamvakou2016depth} 
study how users utilize the  various features 
and functions of \github.
Several studies
\cite{howison2004perils,rainer2005evaluating,treude2018unusual} 
discuss the challenges of mining   
software archives, like \textit{SourceForge} and \github,
arguing that more information is required to make assertions about
users and software projects.


\textbf{c. Databases of malware source code:}
At the time of writing this paper, there are few malware source
code databases and are rarely updated such as project 
\textit{theZoo}~\cite{thezoo}. To the best of our knowledge, there 
does not exist an active archive of malware source code, where 
malware research community can get an enough number of source code to analyze. 

\textbf{d. Database of malware binaries:}
There are well established malware binary collection initiatives, 
such as Virustotal~\cite{total2019virustotal} which provides analysis
result for a malware binary. There are also community based projects 
such as VirusBay~\cite{virusbay} that serve as malware binary sharing
platform.


\textbf{e. Converting binaries to source code:}
A complementary approach is to try to generate the source code
from the binary, but this is a very hard task.
Some works~\cite{vdurfina2011design,vdurfina2013psybot} focus on 
reverse engineering of the malware binary to a high-level language representation, but not source code.
Some other efforts 
 \cite{healey2019source, schulte2018evolving,chen11novel} introduce binary decompilation into readable source code.
However, malware authors use sophisticated obfuscation techniques \cite{saidi2010experiences}\cite{yakdan2016helping,chen2013refined} to make it difficult to reverse engineer a binary into source code.

{\bf f. Measuring and modeling hacking activity.}
Some other studies analyze the underground black market of hacking activities but their starting
point is security forums~\cite{Portnoff2017, Deb2018_USC3,Sapienza2018_USC2}, and as such they
study the dynamics of that community but without retrieving any malware code.



\section{Conclusion}
\vspace{-0.1in}

Our work capitalizes on a great missed opportunity: there are thousands of malware source code repositories on \github.
At the same time, there is a scarcity of 
malware source code, which  is necessary for
certain research studies.

 Our work is arguably the first to develop a systematic approach to extract malware source-code repositories
 at scale from \github.
Our work provides two main tangible outcomes:
(a) we develop \algo, which identifies malware repositories
 with \precision precision, and
(b) we create, possibly, the largest non-commercial malware source code archive with \sourceNum repositories.
Our large scale study provide some interesting trends for both the malware repositories and the dynamics of the malware authors.


We intend to open-source both \algo and the database of malware source code to maximize the impact of our work.
Our ambitious vision is to become the authoritative
source for malware source code  for the research community by providing tools, databases, and benchmarks.

{ \bibliographystyle{plain}
\bibliography{ROKON}
}

\end{document}